\documentclass[11pt,a4paper]{article}

\usepackage{amsfonts}
\begin{document}
\vspace{1.truecm}
\begin{center}

{\LARGE{\bf{ Exact solution of a supersymmetric Gaudin model}}}

\vskip1cm
         
{\large{\sc Julia Breiderhoff}}\vspace{.1cm}\\
{\it Dipartimento di Fisica,Universit\`a di Roma  TRE, }\\
{\it Via Vasca Navale 84, 00146 Roma , Italy}
\vskip0.3cm

{\large{\sc Fabio Musso}}\vspace{.1cm}\\
{\it S.I.S.S.A.}\\
{\it Via Beirut 2/4, I-34013 Trieste, Italy}
\vskip0.3cm

{\large{\sc Orlando Ragnisco}}\vspace{.1cm}\\
{\it  Dipartimento di Fisica,Universit\`a di Roma  TRE,}\\ 
{\it Via Vasca Navale 84, 00146 Roma, Italy} \\
{\it  I.N.F.N. - Sezione di Roma TRE, Roma, Italy}
\vspace{.3cm}
\end{center}
\begin{abstract}
A special case of the  Gaudin model related to the superalgebra $osp(1,2)$ is investigated. An exact solution  in the spin-$\frac{1}{2}$ representation is presented. A complete set of commuting observables is diagonalized, and the corresponding eigenvectors and eigenvalues are explicitly calculated. The approach used in this paper is based on the co-algebra symmetry of the model, already known from the spin-$\frac{1}{2}$ Calogero Gaudin system.
\end{abstract}
\bigskip

\renewcommand{\thesection}{\Roman{section}}
\section{Introduction}
\setcounter{equation}{0}
\renewcommand{\thesection}{\arabic{section}}

The Gaudin model (sometimes called Gaudin magnet) ~\cite{gaudin1}~\cite{gaudin2}~, introduced by M.Gaudin in 1976, is a quantum mechanical system 
involving long-range spin-spin interaction, described by a set of $N$ hamiltonians 
$\mathcal{H}_i$ depending on a set of $N$ arbitrary parameters $\epsilon_i$:
\begin{equation}
\mathcal{H}_i=\sum_{j=1, j\neq i}^N \frac{\sigma^x_i \sigma^x_j+\sigma^y_i \sigma^y_j
+\sigma^z_i \sigma^z_j}{\epsilon_i-\epsilon_j} 
\quad j=1,\dots,N \label{Hi} 
\end{equation}
Out of the operators (\ref{Hi}) Gaudin constructed the further operator 
(dependent also on another set of arbitrary parameters $\eta_i$) 
\begin{equation}
{\mathcal{H}}_G=\sum_{i<j} \frac{ \eta_i- \eta_j}{\epsilon_i-\epsilon_j} 
(\sigma^x_i \sigma^x_j+\sigma^y_i \sigma^y_j
+\sigma^z_i \sigma^z_j)=\sum_{i=1}^N \eta_i\mathcal{H}_i  \label{H}
\end{equation}
selecting it as the Hamiltonian of the system: while 
 $\mathcal{H}_i$ describes just the interaction of 
the $i$-th  spin with all the others, ${\mathcal{H}}_G$ takes care of the mutual interaction
among all spins.\\
Gaudin showed that the 
hamiltonians $\mathcal{H}_i\ $ commute with each other and constructed their common 
eigenvectors by the coordinate Bethe-Ansatz. The algebra corresponding to the observables 
of the model is a loop-algebra, the so-called ``rational Gaudin algebra'', whose properties 
underlie the integrability of the system.
Later, this model has been studied also by Sklyanin through the r-matrix approach ~\cite{sklyanin1}~\cite{sklyanin2} and quite recently its intimate connections with the   Richardson's work on ``pairing force'' 
Hamiltonians ~\cite{richardson} and with BCS model of superconductivity has been unveiled ~\cite{sierra}~\cite{roman}.\\
For the special choice $\eta_i=\epsilon_i \ \forall i$,
the Gaudin Hamiltonian (\ref{H}) takes the simpler, parameter independent, form: 
\begin{equation}
{\mathcal{H}}_G=\sum_{i,j} (\sigma^x_i \sigma^x_j+\sigma^y_i \sigma^y_j
+\sigma^z_i \sigma^z_j) \label{Hmf}
\end{equation}
In the following we will always work
with the ``mean-field'' Hamiltonian (\ref{Hmf}) and we will continue to refer to it,
a bit improperly, as ``Gaudin magnet''.  

Using co-algebras, Ballesteros and coll. 
developed an alternative approach, shortly denoted as BMR in the sequel, to construct 
a second family of commuting observables, independent of the Gaudin set 
$\{\mathcal{H}_i\}$, but sharing with it the element 
(\ref{Hmf})~\cite{balrag}~\cite{mura}~\cite{ramu}. In~\cite{ramu} 
Musso and Ragnisco (MR in the sequel) performed an exact diagonalization  of this alternative family of commuting observables, in the simplest 
finite-dimensional representation, namely the spin-$\frac{1}{2}$ one; their results  
will be tersely recalled  in the second section of this paper. 
We stress that the  existence of two independent complete sets of commuting observables implies 
that the Hamiltonian (\ref{Hmf}) is maximally super-integrable.\\
The superalgebra extension of the notion of Gaudin algebra, and of the related $r-$matrix structure, has been worked out by Macfarlane, Kulish and others in some remarkable papers (see for instance ~\cite{macfarlane}~\cite{kulish}): in particular, they have investigated the Gaudin magnet related to orthosymplectic Lie superalgebra $osp(1,2)$,  and have   obtained for it the complete spectrum  by a suitable generalization of the Bethe-Ansatz.\\
In the present paper, our purpose is  to extend BMR approach to the supersymmetric Gaudin magnet associated with the $osp(1,2)$ superalgebra and to find a complete family of commuting observables independent of the set found by the previous authors in the context of the Bethe-Ansatz. These new observables are in fact a sequence of coproducts of the quadratic Casimir of the algebra $osp(1,2)$. By diagonalizing them, we will construct an exact solution of the model, providing an explicit expression for the associated eigenvectors and eigenvalues. Consequently the Gaudin magnet related to the superalgebra $osp(1,2)$ is still maximally superintegrable.\\
The paper is organized as follows. In section II, we recall the  MR model in the nonsupersymmetric spin-$\frac{1}{2}$ case. In section III we introduce the superalgebra $osp(1,2)$ with its five generators (three bosonic ($E^{\pm},H$) and two fermionic ($F^{\pm}$) ones).  Later,  we recall the notion of graded tensor-product multiplication and show that the coproduct is a homomorphism of $osp(1,2)$. Then, we show that it is possible to derive a completely integrable supersymmetric $N$-body Hamiltonian system in the same way as in the case $sl(2)$, using the properties of co-algebras~\cite{balrag}~\cite{balcorag}. In section IV we construct the common eigenstates of the observables defined in section III. This construction is done in two steps. First, starting from the ``total pseudovacuum'', which is the tensor-product of  the lowest weight vectors pertaining to  each single site and therefore is unique, we built up all the states belonging to the kernel of $\Delta^{(N)}(F^-)$. Then we construct all the remaining states by applying $\Delta^{(N)}(F^+)$ on that kernel, obtaining at the end $3^N$ eigenstates. We notice that all the eigenvectors can be found by applying only the fermionic operator $F^+$, because there is a clear relationship between the fermionic and the bosonic operators: ($F^{\pm})^2=\pm E^{\pm}$). This means that the algebra can actually be generated only by the three operators $F^{\pm},H$. In section V we find the eigenvalues associated to the partial Casimir operators of the system $\Delta^{(h)}(C)$, $(h=2,\dots ,N)$ (coproducts of the Casimir). In section VI we briefly introduce a more general Hamiltonian, corresponding to the so-called $t-J$ models~\cite{gould}, which commutes with the complete family of observables defined in section III and is therefore still completely integrable. The proof that our set of eigenstates is complete is given in the Appendix, where the explicit construction for the $N~=~3$ is also presented.
   
\renewcommand{\thesection}{\Roman{section}}
\section{The MR  model in the nonsupersymmetric spin $\frac{1}{2}$ case}
\renewcommand{\thesection}{\arabic{section}}

In~\cite{ramu} it has been studied a particular, finite-dimensional, $sl(2)$ representation, namely the spin-$\frac{1}{2}$ representation with generators: $\sigma^3$,$\sigma^+$ and $\sigma^-$.\\
With this choice, the $N$th co-product of the Casimir gives us the Hamiltonian of the so-called `Gaudin magnet': 
\begin{eqnarray}
C_N=\Delta^{(N)}(C)&=&[\Delta^{(N)}(\sigma^3)]^2 + 2\{ \Delta^{(N)}(\sigma^+),
\Delta^{(N)}(\sigma^-) \}= \nonumber\\
&=&\sum_{j,k=1}^N(\sigma_j^x \sigma_k^x +\sigma_j^y \sigma_k^y +\sigma_j^z
\sigma_k^z)\equiv {\mathcal{H}}_{G} \label{conocasi}
\end{eqnarray}
The operators $\Delta^{(h)}$, $h=3,\dots ,N$ are defined through the recursive formula ~\cite{balrag}:
\begin{equation}
\Delta^{(h)}=(\Delta^{(h-1)} \otimes id) \Delta^{(2)} \qquad h=3,\dots ,N 
\label{deltah} 
\end{equation}
where $\Delta^{(2)}$ denotes the coproduct associated to the 
usual Hopf-algebra structure defined on $U(g)$, the  universal enveloping 
algebra of a given  Lie algebra $g$, namely:
\begin{equation}
\Delta^{(2)}(X) = X\otimes 1 + 1\otimes X \quad \forall \ X \in g \label{coproduct}
\end{equation}
The quantum system described by the Gaudin Hamiltonian  $H_G$ (\ref{conocasi}) is completely integrable (actually, super-integrable) and the $N-1$ observables commuting among themselves and with the Hamiltonian are provided by the coproducts of the Casimir operator of the algebra $sl(2)$ plus the $N$th coproduct of the operator $\sigma_3$~\cite{balrag},~\cite{balcorag}:
\begin{equation}
\{ \Delta^{(N)} (\sigma^3), C_{2},\dots, C_{N} \} \label{nofamily}
\end{equation}
where we posed for shortness $C_{h}=\Delta^{(h)}(C)$, $h=2,\dots ,N$.
\\
Incidentally, we notice  that in (\ref{conocasi}) we passed from the Lie algebra $sl(2)$ to the Lie algebra $su(2)$, which can be done with the one to one mapping:\\$\sigma_{\pm}=\frac{\sigma_x\pm i\sigma_y}{2};\qquad \sigma_z\equiv \sigma_3$.\\
In~\cite{ramu} the problem of finding the common eigenvectors and the eigenvalues of the set (\ref{nofamily}) has been completely solved.
 
\renewcommand{\thesection}{\Roman{section}}
\section{The Gaudin magnet related to the superalgebra $osp(1,2)$}
\setcounter{equation}{0}
\renewcommand{\thesection}{\arabic{section}}

The simplest superalgebra that generalizes $sl(2)$  is the orthosymplectic superalgebra $osp(1,2)$ defined by commutation and anticommutation rules (supercommutation rules). As it is well known, the basic  difference between the nonsupersymmetric case and the supersymmetric case is the presence of a grading, which classifies the elements of the algebra in two families, the bosons with even degree and the fermions with odd degree. These degrees define the supercommutation rules. So we have to revise the technique we used in the case of the algebra $sl(2)$ according to the above considerations. We have to introduce a graded tensor-product multiplication and to verify that the coproduct is still a homomorphism of the superalgebra under scrutiny. The superalgebra $osp(1,2)$ has five generators, three bosonic ones: $H,E^+,E^-$ and two fermionic ones: $F^+,F^-$. The relations defining the superalgebra are the following ~\cite{frappat}:
\begin{eqnarray}
&[H,E^\pm]_{-} = \pm 2E^\pm &\qquad   [E^+,E^-]_{-} = H \label{commo1}\\
&[H,F^\pm]_{-} = \pm F^\pm &\qquad  [F^+,F^-]_{+} = H\label{commo2} \\ 
&[E^\pm,F^\mp]_{-} = -F^\pm &\qquad  [F^\pm,F^\pm]_{+} = \pm2E^\pm 
\label{commo3} 
\end{eqnarray}
We immediately see that the operators $H,E^+,E^-$ generate the $sl(2)$ Lie algebra. Moreover, as the anticommutation rules entail $(F^\pm)^2~=~E^\pm$, some authors (see for instance \cite{kulish}) define $osp(1,2)$ only in terms of the three generators $H,~F^+,~F^-$ .\\
The Casimir of this algebra is:
\begin{equation}
\label{casimir}
C = H^2 + 2 (E^+E^- + E^-E^+) - (F^+F^- - F^-F^+)
\end{equation}
As we shall see in the sequel, the $N$th coproduct of Casimir gives us the Hamiltonian of the `supersymmetric Gaudin magnet':
\begin{eqnarray}
 C_N& =& \Delta^{(N)}(C) = [\Delta^{(N)}(H)]^2 + 2[\Delta^{(N)}(E^+),\Delta^{(N)}(E^-)]_+ -\nonumber\\
&&{}- [\Delta^{(N)}(F^+),\Delta^{(N)}(F^-)]_-=\nonumber\\
{}&&=\sum_{j\neq i}^N\mathbf{S}^i \mathbf{S}^j-F_+^iF_-^j+F_-^iF_+^j+2Nj(j+\frac{1}{2})I\equiv \mathcal{H}_{G_s}\label{hgi}
\end{eqnarray}
where $I$ is the identity operator, the (positive) integer or half-integer $j$ denotes the spin in each site (\footnote{ A further generalization is obtained by allowing site-dependent representations $(j_1, \dots , j_N)$: of course in this case the constant $2Nj(j+\frac{1}{2})$ is then replaced by $2\sum_{k=1}^Nj_k(j_k+\frac{1}{2})$.}) and ${\bf{S}}^l\equiv (H^l,E_+^l+E_-^l,-i(E_+^l-E_-^l)).$\\           
Again we pose $C_h=\Delta^{(h)}(C)$, $h=2,\dots ,N$; in this representation a complete set of independent commuting observables is provided by 
\begin{equation}
\{\Delta^{(N)}(H),C_2,\dots ,C_N\} \label{osser}\end{equation}
\renewcommand{\thesection}{\Roman{section}}  
\subsection{Coproduct approach for the superalgebra $osp(1,2)$}
\renewcommand{\thesection}{\arabic{section}}

As it has been seen in~\cite {balrag}, for the  construction of a $N$-particle completely integrable Hamiltonian system the coproduct has to be a coassociative homomorphism of the algebra under scrutiny, so that the coproducts of the generators realize the same algebra as the original generators. Our aim is to construct a $N$-particle completely integrable supersymmetric Hamiltonian system, using the same technique described in~\cite{balrag},~\cite{ramu}. It is then necessary to demonstrate that the coproduct is a coassociative homomorphism for the superalgebra $S\equiv osp(1,2)$ as well, i.e.  it must hold:
 \begin{eqnarray}
[\Delta(a),\Delta(b)]_{S\otimes S}& =& \Delta ([a,b]_S)\qquad \forall a,b \in S\\
(\Delta \otimes id) \Delta &=& (id \otimes \Delta) \Delta
\end{eqnarray}
where by $[\, ,\,]$ we have denoted the supercommutator.\\
The above formulas involve tensor-products multiplication,  which has to be defined in such a way that associativity is guaranteed. In fact, associativity of  tensor-product multiplication entails  coassociativity of the coproduct.
\renewcommand{\thesection}{\Roman{section}}
\subsubsection{Tensor-product}
\renewcommand{\thesection}{\arabic{section}}
We know from~\cite{frappat} that for binary tensor products an associative, necessarily graded, multiplication is given by: 
\begin{equation}
(a \otimes b) (c \otimes d ) = ac \otimes bd (-1)^{deg (b) \dot deg (c)}\label{grading}\label{grad2}
\end{equation}
Defining the degree of binary  tensor-product in the natural way:
\begin{equation}
deg(a\otimes b)=deg(a)+deg(b)\qquad mod2 \label{mod2}
\end{equation}
we obtain for  three element tensor-products the following equations:
\begin{eqnarray}
&&((a\otimes b)\otimes c)((d\otimes e)\otimes f)=(a\otimes b)(d\otimes e)\otimes cf (-1)^{deg(c)deg(d\otimes e)}=\nonumber\\
&&=ad\otimes be\otimes cf (-1)^{deg(b)deg(d)+deg(c)deg(d)+deg(c)deg(e)}\label{first}
\end{eqnarray}
and
\begin{eqnarray}
&&(a\otimes (b\otimes c))(d\otimes (e\otimes f))=ad\otimes (b\otimes c)(e\otimes f) (-1)^{deg(b\otimes c)deg(d)}=\nonumber\\
&&=ad\otimes be\otimes cf (-1)^{deg(b)deg(d)+deg(c)deg(d)+deg(c)deg(e)}\label{second}
\end{eqnarray}
As (\ref{first}) and (\ref{second}) coincide, the multiplication between three element tensor-products  is associative.
We can extend this result to the general case of multiplication between $N$ element tensor-products obtaining:
\begin{eqnarray}
&&(a_1\otimes a_2 \otimes \dots \otimes a_N)(b_1\otimes b_2 \otimes \dots \otimes b_N) =\nonumber\\
&&(-1)^{\sum_{i<j=2}^N deg(a_j)deg(b_i)}(a_1b_1\otimes a_2b_2 \otimes \dots \otimes a_Nb_N) \qquad \forall a_j,b_i \in \mathcal{S}\nonumber\\
&&{}
\end{eqnarray}
\renewcommand{\thesection}{\Roman{section}}
\subsection{Construction of integrable systems of $N$ particles}
\renewcommand{\thesection}{\arabic{section}}
Generalizing (\ref{mod2}), we define the degree of the $N$ element tensor-product as:
\begin{equation}
deg(a_1\otimes a_2\otimes \dots \otimes a_N)=deg(a_1)+deg(a_2)+\dots +deg(a_N)\qquad mod2 \label{ntensor}
\end{equation}
From this definition it follows straight away that the coproduct $\Delta^{(N)}(X_i)$ preserves the degree of the generic operator $X_i$. In fact we have:
\begin{equation}
\Delta^{(N)}(X_i)=X_i\otimes \overbrace{1\otimes \dots \otimes 1}^{N-1}+1\otimes X_i\otimes \overbrace{1\otimes \dots \otimes 1}^{N-2}+\dots + \overbrace{1\otimes \dots \otimes 1}^{N-1}\otimes X_i
\end{equation}
Hence:
\begin{equation}
deg(\Delta^{(N)}(X_i))=deg(X_i)+(N-1)deg(1)
\end{equation}
But $deg(1)=0$,  so we obtain:
\begin{equation}
deg(\Delta^{(N)}(X_i))=deg(X_i)
\end{equation}
We can immediately see that, as in the case $sl(2)$, a completely integrable supersymmetric  
$N$-body Hamiltonian system can be constructed through 
the BMR approach which yields the complete family of commuting observables (\ref{osser}). 

\renewcommand{\thesection}{\Roman{section}}
\section{Construction of the simultaneous eigenstates in the case $osp(1,2)$}
\setcounter{equation}{0}
\renewcommand{\thesection}{\arabic{section}}

In the sequel, we will restrict our consideration to the case with spin $j$ in each site. Our aim is to find the common eigenstates of the observables (\ref{osser}), which form a basis for the Hilbert-space of the problem. As reference state, the so-called `` pseudovacuum'', we take   the tensor-product of the states annihilated by the single-particle  operators $F_i^-$, namely the tensor-product of the $N$ single-particle lowest weight vectors ${|0 \rangle}_i$, $i=1,\dots,N$; as in the non supersymmetric case, we denote it as $\Psi (0,0)$.\\
We will show that  the common eigenstates of the family of observables (\ref{osser}) take the form:
\begin{equation}
\Phi(k;m_l,s_{m_l};\dots;0,0)=[\Delta^{(N)}(F^+)]^{k-m_l} \Psi (m_l,s_{m_l};\dots ;0,0)
\end{equation}
where $\Psi (m_l,s_{m_l};\dots ;0,0)$ is an element of the basis spanning the kernel $ Ker (\Delta^{(s_{m_l})}(F^- ))$, obtained through the recursive formula:
\begin{eqnarray} 
&&\Psi (m_l,s_{m_l};\dots ;0,0) = \nonumber \\
{}&&= \sum_{i=0}^{m_l-m_{l-1}} a_i \Delta^{(s_{m_l}-1)}(F^+)^{m_l-m_{l-1}-i}(F_{s_{m_l}}^+)^i \Psi (m_{l-1},s_{m_{l-1}};\dots ;0,0)\nonumber\\
{}&&m_l = 1,2,\dots,N \qquad  s_{m_l} = 2,3,\dots,N\qquad l=1,2\dots N-1 \nonumber\\
{}&&m_{r-1} < m_r\qquad s_{m_{r-1}} < s_{m_r}\qquad r = 1,2,\dots,l \label{ricur}
\end{eqnarray}
In (\ref{ricur}) $s_{m_l}$ indicates the number of sites involved; $m_l$ indicates the total excitation of the system, namely $m_l -N$ is  the eigenvalue of $\Delta^{(N)}(H)$.\\
$F_{s_{m_l}}^+$ is a shorthand for $\overbrace{ id \otimes \dots \otimes id }^{s_{m_l}-1} \otimes F^+ \otimes \overbrace{ id \otimes \dots \otimes id}^{N-s_{m_l}}$ and the coefficients $a_i$ are determined by the equation
\begin{equation}
\Delta^{(s_{m_l})}(F^-) \Psi (m_l,s_{m_l};\dots;0,0) = 0\label{kernel}
\end{equation}
We also  define $\Delta ^{(1)}(F^{\pm}) =F^{\pm}$.\\
The following condition is satisfied:
\begin{equation}
4j\geq m_l-m_{l-1} \geq 1  \label{dis}
\end{equation}
Plugging (\ref{ricur}) in (\ref{kernel}) we obtain an equation which determines $a_i$.\\
To determine the states $\Psi(m_l,s_{m_l};\dots ;0,0)$ we will proceed in the following way:
\begin{itemize}
\item {\bf{1)}} First, we show that the states $\Psi(m_l,s_{m_l};\dots ;0,0)$ are eigenstates of the operator $\Delta^{(s_{m_l})}(H)$ with eigenvalues $(m_l-s_{m_l})$. This implies that  these states are also eigenvectors of the operators $\Delta^{(s_{m_l})}(C)$.
\item {\bf{2)}} Next, we calculate the coefficients $a_i$ through  a recursive formula.
\item {\bf{3)}} Finally, we discuss  the special  case $j=\frac{1}{2}$ and fully characterize the quantum numbers labelling the states.
\end{itemize}
\renewcommand{\thesection}{\Roman{section}}
\subsection{Calculating the eigenvalues of $\Delta^{(s_{m_l})}(H)$}
\renewcommand{\thesection}{\arabic{section}}
Writing:
\begin{equation}
\Delta^{(s)} (H)=\Delta^{(s-1)} (H) \otimes id + \Delta^{(s-1)} (id) 
\otimes H\label{delt}
\end{equation}
and using the formula 
\begin{equation}
[ H,(F^+)^i ]_{-} = i (F^+)^i\label{effe}
\end{equation}
obtained by induction from the relation (\ref{commo2}) $[H,F^+]_-=F^+$, we obtain:
\begin{eqnarray}
&&\Delta^{(s_{m_l})}(H) \Psi (m_{l},s_{m_l};\dots;0,0)=\nonumber\\
{}&&=(\Delta^{(s_{m_{l}}-1)} (H) \otimes id + \Delta^{(s_{m_l}-1)} (id)
\otimes H) \Psi (m_{l},s_{m_l};\dots;0,0) =\nonumber\\
{}&&= \sum_{i=0}^{m_{l}-m_{l-1}} a_i  \Bigg[ \bigg( \Delta^{(s_{m_l}-1)}(H(F^+)^{m_l-m_{l-1}-i})(F_{s_{m_l}}^+)^i \bigg)+\nonumber\\
{}&&+\bigg( \Delta^{(s_{m_l}-1)}(F^+)^{m_l-m_{l-1}-i}H_{s_{m_l}}(F_{s_{m_l}}^+)^i 
\bigg) \Bigg] \Psi (m_{{l}-1},s_{m_{{l}-1}};\dots;0,0)=\nonumber\\
{}&&= \sum_{i=0}^{m_{l}-m_{l-1}} a_i \Bigg[ \Delta^{(s_{m_l}-1)} \big((F^+)^{m_l-m_{l-1}-i} H \big) (F_{s_{m_l}}^+)^i) +\nonumber\\
{}&&\Delta^{(s_{m_l}-1)}(F^+)^{m_l-m_{l-1}-i}(F_{s_{m_l}}^+)^i H_{s_{m_l}} + (m_l-m_{l-1}-i)\Delta^{(s_{m_l}-1)}(F^+)^{m_l-m_{l-1}-i} (F_{s_{m_l}}^+)^i +\nonumber\\
{}&&+ {i}\Delta^{(s_{m_l}-1)}(F^+)^{m_l-m_{l-1}-i}(F_{s_{m_l}}^+)^i \Bigg] \Psi (m_{{l}-1},s_{m_{{l}-1}};\dots;0,0)=\nonumber\\
{}&&= \sum_{i=0}^{m_{l}-m_{l-1}} a_i \Bigg[ \Delta^{(s_{m_l}-1)}(F^+)^{m_l-m_{l-1}-i}(F_{s_{m_l}}^+)^i \Big( \Delta^{(s_{m_l}-1)}(H) + H_{s_{m_l}} \Big) +\nonumber\\
{}&&+(m_{l}-m_{l-1})\Delta^{(s_{m_l}-1)}(F^+)^{m_l-m_{l-1}-i} (F_{s_{m_l}}^+)^i \Bigg] \Psi (m_{{l}-1},s_{m_{{l}-1}};\dots;0,0)=\nonumber\\
{}&&= \sum_{i=0}^{m_{l}-m_{l-1}} a_i \Bigg[ \Delta^{(s_{m_l}-1)}(F^+)^{m_l-m_{l-1}-i}( m_{l-1}-s_{m_{l}} )(F_{s_{m_l}}^+)^i +\nonumber\\
{}&&+( m_l-m_{l-1} ) \Delta^{(s_{m_l}-1)}(F^+)^{m_l-m_{l-1}-i} (F_{s_{m_l}}^+)^i \Bigg] \Psi (m_{{l}-1},s_{m_{{l}-1}};\dots;0,0)=\nonumber\\
{}&&= \sum_{i=0}^{m_{l}-m_{l-1}} a_i \bigg( (m_{l}-s_{m_{l}})  \Delta^{(s_{m_l}-1)}(F^+)^{m_l-m_{l-1}-i} (F_{s_{m_l}}^+)^i \bigg) \Psi (m_{{l}-1},s_{m_{{l}-1}};\dots;0,0)=\nonumber\\
{}&&(m_l-s_{m_l}) \sum_{i=0}^{m_{l}-m_{l-1}} a_i \bigg( \Delta^{(s_{m_l}-1)}(F^+)^{m_l-m_{l-1}-i} (F_{s_{m_l}}^+)^i \bigg) \Psi (m_{{l}-1},s_{m_{{l}-1}};\dots;0,0)
\end{eqnarray}
Hence:
\begin{equation}
\Delta^{(s_{m_{l}})} (H) \Psi (m_{l},s_{m_{l}};\dots;0,0)
= (m_{l} - s_{m_{l}}) \Psi (m_{l},s_{m_{l}};\dots;0,0) \label{deltah1}
\end{equation}
Using commutation relations (\ref{commo1}):
\begin{eqnarray}
&& \Delta^{(s_{m_{l}})} (H) \Psi (m_{l},s_{m_{l}};\dots;0,0)=\nonumber\\
{}&& =[\Delta^{(s_{m_{l}})} (E^+),\Delta^{(s_{m_{l}})} (E^-)]_-
\Psi (m_{l},s_{m_{l}};\dots;0,0)= \nonumber\\
{}&&=  -\Delta^{(s_{m_{l}})} (E^-)\Delta^{(s_{m_{l}})} (E^+)
\Psi (m_{l},s_{m_{l}};\dots;0,0)\label{deltah2}
\end{eqnarray}
Being $\Delta^{(s_{m_{l}})} (E^-)\Delta^{(s_{m_{l}})}(E^+)$ semi-positive, it has to be:
\begin{equation}
\quad s_{m_{l}} - m_{l} \geq 0 \quad  
\label{essemme}
\end{equation}
To derive (\ref{deltah2}) we used the relation $E^\pm = {(F^\pm)}^2$, following from (\ref{commo3}), and the coassociativity of the superalgebra coproduct, yielding:
\begin{equation}
\Delta^{(s_{m_l})}(E^\pm) = \Delta^{(s_{m_l})}(F^\pm) \Delta^{(s_{m_l})}(F^\pm)
\end{equation}
Hence due to (\ref{kernel}) we have:
\begin{eqnarray} 
&&\Delta^{(s_{m_l})}(E^-) \Psi (m_{l},s_{m_l};\dots;0,0) = \nonumber\\
&&= \Delta^{(s_{m_l})}(F^-) \Delta^{(s_{m_l})}(F^-)\Psi (m_{l},s_{m_l};\dots;0,0) = 0
\end{eqnarray}
\renewcommand{\thesection}{\Roman{section}}
\subsection{Calculating the coefficients $a_i$}
\renewcommand{\thesection}{\arabic{section}}
Using (\ref{delt}) and the following formula obtained from (\ref{commo2}) and (\ref{effe}):
\begin{equation}
F^-(F^+)^k = (-1)^k(F^+)^kF^- + (F^+)^{k-1} \Big( \bigg[ \frac{k}{2} \bigg] + \big(\frac{1}{2}(1-(-1)^k)H \big) \Big)
\end{equation}
we obtain:
\begin{eqnarray}
&&\Delta^{(s_{m_l})}(F^-) \Psi (m_l,s_{m_l};\dots;0,0) = \nonumber\\ 
&&= \Big( \Delta^{(s_{m_{l}}-1)} (F^-) + F_{s_{m_l}}^- \Big) \Big( \sum_{i=0}^{m_{l}-m_{l-1}} a_i \Delta^{(s_{m_l}-1)}(F^+)^{m_l-m_{l-1}-i}(F_{s_{m_l}}^+)^i \Big) \times \nonumber\\
&&\times \Psi (m_{l-1},s_{m_{l-1}};\dots;0,0)= \nonumber\\
&&= \sum_{i=0}^{m_{l}-m_{l-1}} a_i \Bigg[ (-1)^{m_l-m_{l-1}} \Delta^{(s_{m_l}-1)}(F^+)^{m_l-m_{l-1}-i} (F_{s_{m_l}}^+)^i \Delta^{(s_{m_{l}}-1)} (F^-)+ \nonumber\\
&&+ \Delta^{(s_{m_l}-1)}(F^+)^{m_l-m_{l-1}-i-1}(F_{s_{m_l}}^+)^i \Big( \bigg[ \frac{m_l-m_{l-1}-i}{2} \bigg] + \nonumber \\
&&+ \frac{1}{2} \Big( 1-(-1)^{m_l-m_{l-1}-i} \Big) \Delta^{(s_{m_{l}}-1)} (H)\Big)+ \nonumber\\
&&+ (-1)^{m_l-m_{l-1}-i} \Delta^{(s_{m_l}-1)}(F^+)^{m_l-m_{l-1}-i} \bigg( (-1)^i(F_{s_{m_l}}^+)^i  F_{s_{m_l}}^- + \nonumber\\
&&+ (F_{s_{m_l}}^+)^{i-1}\Big( \Big[ \frac{i}{2} \Big] + \frac{1}{2} (1-(-1)^i)H_{s_{m_l}}\bigg) \Bigg] \Psi (m_{l-1},s_{m_{l-1}};\dots;0,0) = \nonumber\\
&&= \sum_{i=0}^{m_{l}-m_{l-1}} a_i \Bigg[  \Delta^{(s_{m_l}-1)}(F^+)^{m_l-m_{l-1}-i-1}(F_{s_{m_l}}^+)^i \Big( \bigg[ \frac{m_l-m_{l-1}-i}{2} \bigg] + \nonumber\\
&&+ \frac{1}{2} \Big( 1-(-1)^{m_l-m_{l-1}-i} \Big) ( m_{l-1} - s_{m_l} + 1) \Big)+ \nonumber\\
&&+ (-1)^{m_l-m_{l-1}-i} \Delta^{(s_{m_l}-1)}(F^+)^{m_l-m_{l-1}-i} (F_{s_{m_l}}^+)^{i-1}\Big( \Big[ \frac{i}{2} \Big] - \nonumber\\
&&- \frac{1}{2} (1-(-1)^i)\Big) \Bigg] \Psi (m_{l-1},s_{m_{l-1}};\dots;0,0).\label{equfminus}   
\end{eqnarray}
To derive (\ref{equfminus}), we used the property:
\begin{equation}
\Delta^{(s_{m_l}-1)}(F^-)\Psi (m_{l-1},s_{m_{l-1}};\dots;0,0)= 0
\end{equation}
Also, we notice that, being $s_{m_l}-1 < s_{m_l}$  it holds:
\begin{equation}
F_{s_{m_l}}^- \Psi (m_{l-1},s_{m_{l-1}};\dots;0,0)= 0
\end{equation}
We go on calculating the coefficients $a_i$:
\begin{eqnarray}
&&\Delta^{(s_{m_l})}(F^-) \Psi (m_l,s_{m_l};\dots;0,0) =\\ 
&&=\Bigg[ \sum_{i=0}^{m_{l}-m_{l-1}-1} a_i  \Delta^{(s_{m_l}-1)}(F^+)^{m_l-m_{l-1}-1-i}(F_{s_{m_l}}^+)^i \times \nonumber\\
&&\times  \Big( \bigg[ \frac{m_l-m_{l-1}-i}{2} \bigg] + \frac{1}{2} \Big( 1-(-1)^{m_l-m_{l-1}-i} \Big) ( m_{l-1} - s_{m_l} + 1) \Big) +\nonumber\\
&&+ \sum_{i=1}^{m_{l}-m_{l-1}} a_i (-1)^{m_l-m_{l-1}-i} \Delta^{(s_{m_l}-1)}(F^+)^{m_l-m_{l-1}-i}(F_{s_{m_l}}^+)^{i-1} \times \nonumber\\
&&\times \Big( \Big[ \frac{i}{2} \Big] - \frac{1}{2} (1-(-1)^i)\Big) \Bigg] \Psi (m_{l-1},s_{m_{l-1}};\dots;0,0) =\nonumber\\
&&= \sum_{i=0}^{m_{l}-m_{l-1}-1}  \Delta^{(s_{m_l}-1)}(F^+)^{m_l-m_{l-1}-1-i}(F_{s_{m_l}}^+)^i \times\nonumber \\
&&\times \bigg[ a_i \Big( \bigg[ \frac{m_l-m_{l-1}-i}{2} \bigg] + \frac{1}{2} \Big( 1-(-1)^{m_l-m_{l-1}-i} \Big) ( m_{l-1} - s_{m_l} + 1) \Big) +\nonumber\\
&&+ a_{i+1} (-1)^{m_l-m_{l-1}-(i+1)} \Big( \Big[ \frac{i+1}{2} \Big] - \frac{1}{2} (1-(-1)^i+1)\Big) \bigg] \times \nonumber\\
&&\times \Psi (m_{l-1},s_{m_{l-1}};\dots;0,0)\label{fmino}  
\end{eqnarray}
The previous expression will vanish iff the  coefficients $a_i$ obey the recursive formula:
\begin{equation}
a_{i+1}=\\
a_i (-1)^{m_l-m_{l-1}-i} \frac{ \bigg[ \frac{m_l-m_{l-1}-i}{2} \bigg] + \frac{1}{2} \Big( 1-(-1)^{m_l-m_{l-1}-i} \Big) ( m_{l-1} - s_{m_l} + 1)}{ \Big[ \frac{i+1}{2} \Big] - \frac{1}{2} (1-(-1)^i+1)}
\end{equation}
\renewcommand{\thesection}{\Roman{section}}
\subsection{Restrictions on $k$ and the special case $j=\frac{1}{2}$}
\renewcommand{\thesection}{\arabic{section}}
We notice that it must hold $k\geq m_l$. Moreover, as the state $\Phi (k;m_l,s_{m_l};\dots ;0,0 )$ has spin $k - N$, it also has to be $k \leq 2N$.\\
This condition can be further restricted imposing spin-flip symmetry, entailing: $k \leq 2N-m_l$.\\
\\
In the special case $j=\frac{1}{2}$, $ m_l - m_{l-1}$ can take two values: 1 or 2. So we have two cases:\\
{\bf{Case $1)\qquad m_l-m_{l-1}=1\qquad \rightarrow i=0,1$
\begin{equation}
a_{i+1} = a_i (-1)^{1-i} \frac{ \bigg[ \frac{1-i}{2} \bigg] + \frac{1}{2} \Big( 1-(-1)^{1-i} \Big) ( m_{l-1} - s_{m_l} + 1)}{ \Big[ \frac{i+1}{2} \Big] - \frac{1}{2} (1-(-1)^i+1)}
\end{equation}
\begin{equation}
a_1 = a_0 ( m_{l-1} - s_{m_l} + 1)\label{condone}
\end{equation}
Case $2)\qquad m_l-m_{l-1}=2 \qquad \rightarrow i = 0,1,2$
\begin{equation}
a_{i+1} = a_i (-1)^{i} \frac{ \bigg[ \frac{2-i}{2} \bigg] + \frac{1}{2} \Big( 1-(-1)^{i} \Big) ( m_{l-1} - s_{m_l} + 1)}{ \Big[ \frac{i+1}{2} \Big] - \frac{1}{2} (1-(-1)^i+1)}
\end{equation}
\begin{equation}
a_1 = - a_0 \label{condtwo1}
\end{equation}
\begin{equation}
a_2 = - a_1 ( m_{l-1} - s_{m_l} + 1) = a_0 ( m_{l-1} - s_{m_l} + 1)\label{condtwo2}
\end{equation}}}
In both cases, we have the condition: 
\begin{equation}
s_{m_l}>m_{l-1}+1  \label{ine}
\end{equation}
Indeed, from (\ref{dis}) and (\ref{essemme}) it follows that $s_{m_l} \geq m_{l-1}+1$; however,  
if equality holds  $a_0$ is the only non-vanishing coefficient, and (\ref{equfminus}) becomes:
\begin{equation}
a_0 \Delta^{(s_{m_l}-1)}(F^-F^+) \Psi (m_{{l}-1},s_{m_{{l}-1}};\dots;0,0) \neq 0
\end{equation}
which doesn't verify (\ref{kernel}).\\
We rewrite the simultaneous eigenstates of the set of observables (\ref{osser}):
\begin{equation}
\Phi(k;m_l,s_{m_l};\dots;0,0) = [\Delta^{(N)}(F^+)]^{k-m} \Psi (m_l,s_{m_l};\dots;0,0)\label{phi}
\end{equation}
\begin{eqnarray}
N \geq s_{m_r} \geq m_r &\qquad s_{m_r}>s_{m_{r-1}} &\qquad m_r>m_{r-1}\nonumber\\
{} s_{m_r}>m_{r-1}+1 &\qquad  m_l \leq k \leq 2N-m_l &\qquad  m_r-m_{r-1}=1,2\nonumber\\
{}\label{cond}
\end{eqnarray}
\begin{eqnarray}
&&\Psi (m_l,s_{m_l};\dots;0,0) =\nonumber\\
{}&&= \sum_{i=0}^{m_l-m_{l-1}} a_i \Delta^{(s_{m_l}-1)}(F^+)^{m_l-m_{l-1}-i}(F_{s_{m_l}}^+)^i \Psi (m_{l-1},s_{m_{l-1}};\dots;0,0)\nonumber\\
{}&&\label{states2} 
\end{eqnarray}
To construct explicit formulas for the eigenstates we recall that,
for spin $\frac{1}{2}$, the fundamental representation is $3$-dimensional: 
the $osp(1,2)$ generators read:
\begin{eqnarray}
H =
\left( \begin{array}{ccc}
1 & 0 & 0 \\
0 & -1 & 0 \\
0 & 0 & 0 
\end{array} \right) &\qquad
E^+ =
\left( \begin{array}{ccc}
0 & 1 & 0 \\
0 & 0 & 0 \\
0 & 0 & 0 \end{array} \right) &\qquad
E^- = 
\left( \begin{array}{ccc}
0 & 0 & 0 \\
1 & 0 & 0 \\
0 & 0 & 0 
\end{array} \right)\nonumber\\
F^+ =
\left( \begin{array}{ccc}
0 & 0 & 1 \\
0 & 0 & 0 \\
0 & 1 & 0 
\end{array} \right) &\qquad
F^- =
\left( \begin{array}{ccc}
0 & 0 & 0 \\
0 & 0 & -1 \\
1 & 0 & 0 
\end{array} \right) &\qquad  \label{generators}
\end{eqnarray}
\\
Consequently the vacuum $\Psi (0,0)$ in the spin-$\frac{1}{2}$ representation is:
\begin{equation}
\Psi (0,0) = \overbrace{\bigg( \hspace{-1.2mm}\begin{array}{c} \scriptstyle{0} \vspace{-5mm}\\ \scriptstyle{1}\vspace{-5mm}\\ \scriptstyle{0} \end{array} \hspace{-1.2mm} \bigg)\dots\dots \bigg( \hspace{-1.2mm}\begin{array}{c} \scriptstyle{0} \vspace{-5mm}\\ \scriptstyle{1}\vspace{-5mm}\\ \scriptstyle{0} \end{array} \hspace{-1.2mm} \bigg)}^{N}\label{vacuum}
\end{equation}
Of course, there are  three possible single-particle states,\\
namely: $\bigg( \hspace{-1.2mm}\begin{array}{c} \scriptstyle{0} \vspace{-5mm}\\ \scriptstyle{1}\vspace{-5mm}\\ \scriptstyle{0} \end{array} \hspace{-1.2mm} \bigg)$,$ \bigg( \hspace{-1.2mm}\begin{array}{c} \scriptstyle{0} \vspace{-5mm}\\ \scriptstyle{0}\vspace{-5mm}\\ \scriptstyle{1} \end{array} \hspace{-1.2mm} \bigg)$, $\bigg( \hspace{-1.2mm}\begin{array}{c} \scriptstyle{1} \vspace{-5mm}\\ \scriptstyle{0}\vspace{-5mm}\\ \scriptstyle{0} \end{array} \hspace{-1.2mm} \bigg)$\\
where $ \bigg( \hspace{-1.2mm}\begin{array}{c} \scriptstyle{0} \vspace{-5mm}\\ \scriptstyle{0}\vspace{-5mm}\\ \scriptstyle{1} \end{array} \hspace{-1.2mm} \bigg)=F^+\bigg( \hspace{-1.2mm}\begin{array}{c} \scriptstyle{0} \vspace{-5mm}\\ \scriptstyle{1}\vspace{-5mm}\\ \scriptstyle{0} \end{array} \hspace{-1.2mm} \bigg)$ denotes the fermionic state and  $\bigg( \hspace{-1.2mm}\begin{array}{c} \scriptstyle{1} \vspace{-5mm}\\ \scriptstyle{0}\vspace{-5mm}\\ \scriptstyle{0} \end{array} \hspace{-1.2mm} \bigg)=E^+\bigg( \hspace{-1.2mm}\begin{array}{c} \scriptstyle{0} \vspace{-5mm}\\ \scriptstyle{1}\vspace{-5mm}\\ \scriptstyle{0} \end{array} \hspace{-1.2mm} \bigg)$ the bosonic one.
\renewcommand{\thesection}{\Roman{section}}
\section{Eigenvalues of the partial Casimir operator $C_h$ in the case $osp(1,2)$}
\setcounter{equation}{0}
\renewcommand{\thesection}{\arabic{section}}
We will show that the states (\ref{phi}) are indeed eigenstates of the set of observables (\ref{osser}) and then calculate the eigenvalues of the partial Casimir operator $C_h$. First of all we notice that:
\begin{equation}
\Delta^{(N)}(H) \Phi (k;m_l,s_{m_l};\dots ;0,0) = (k-N) \Phi (k;m_l,s_{m_l};\dots ;0,0) \label{states}
\end{equation}
Now we consider the action of the partial Casimir $C_h$ on the generic state (\ref{phi}). 
Using commutation relations (\ref{commo1},\ref{commo2},\ref{commo3}), we rewrite $C_h$ in the
equivalent form:
\begin{eqnarray*}
&& C_h=[\Delta^{(h)}(H)]^2 + 2 \Delta^{(h)}(E^+) \Delta^{(h)}(E^-)+\\
{}&&+ 2\Delta^{(h)}(E^-) \Delta^{(h)}(E^+)-\Delta^{(h)}(F^+) \Delta^{(h)}(F^-)+ \Delta^{(h)}(F^-) \Delta^{(h)}(F^+)=\\
{}&&= [\Delta^{(h)}(H)]^2 + 4 \Delta^{(h)}(E^+) \Delta^{(h)}(E^-) - \Delta^{(h)}(H)-2 \Delta^{(h)}(F^+) \Delta^{(h)}(F^-)
\end{eqnarray*}
As all $C_h$ commute with the $N$th coproduct of any of the operators (\ref{generators}), we only have to worry about the action of the partial Casimir $C_h=\Delta^{(h)}(C)$, ($h=2,\dots ,N)$ on the states (\ref{states2}):
\begin{eqnarray}
&& C_{h} \Psi ( m_l,s_{m_l};\dots;0,0 )= \nonumber\\
{}&&= \Big( [\Delta^{(h)}(H)]^2 + 4 \Delta^{(h)}(E^+) \Delta^{(h)}(E^-) - \Delta^{(h)}(H)-2 \Delta^{(h)}(F^+) \Delta^{(h)}(F^-) \Big)\times\nonumber\\
{}&&\times \big( \sum_{i=0}^{m_l-m_{l-1}} a_i \Delta^{(s_{m_l}-1)}(F^+)^{m_l-m_{l-1}-i}(F_{s_{m_l}}^+)^i \big)\Psi (m_{l-1},s_{m_{l-1}};\dots ;0,0)\label{eigenvalue1}   
\end{eqnarray}
There are two possibilities: $(i)\, h \geq s_{m_l}$\quad and\quad $ (ii)\, h < s_{m_l}$\\ In case $(i)$, we know that:
$$\Delta^{(h)}(F^-) \Psi (m_l,s_{m_l};\dots ;0,0)=0 \rightarrow \Delta^{(h)}(E^-) \Psi (m_l,s_{m_l};\dots;0,0)=0$$
and using equation (\ref{deltah1}) we obtain:
\begin{equation}
C_{h}\Psi (m_l,s_{m_l};\dots;0,0)=(h-m_{l})(h-(m_{l}-1)) \Psi (m_l,s_{m_l};\dots;0,0)\label{eigenvalue2}
\end{equation}
On the other hand, in case $(ii)$, the Casimir operator $C_{h}$ commutes both with $F_{(s_{m_l})}^+$ and with $\Delta^{(s_{m_l}-1)}(F^+)$, so that we obtain:
\begin{eqnarray}
&& C_{h}\Psi (m_l,s_{m_l};\dots;0,0)=\label{eigenvalue3}\\
&& \big( \sum_{i=0}^{m_l-m_{l-1}} a_i \Delta^{(s_{m_l}-1)}(F^+)^{m_l-m_{l-1}-i}(F_{s_{m_l}}^+)^i\big) C_{h} \Psi (m_{l-1},s_{m_{l-1}};\dots;0,0)\nonumber
\end{eqnarray}
Again we have two possibilities: if  $h \geq s_{m_l-1}$, we have:
\begin{eqnarray}
&& \! \! \! \! \!C_{h} \Psi(m_{l-1},s_{m_{l-1}};\dots;0,0)=\nonumber\\
{}&& \! \! \! \!\! =(h-(m_{l-1}))(h-(m_{l-1}-1))) \Psi (m_{l-1},s_{m_{l-1}};\dots;0,0)
\label{eigenvalue4} 
\end{eqnarray}
otherwise $C_{h}$ will work directly on $\Psi (m_{l-2},s_{m_{l-2}};\dots;0,0)$.\\ We can proceed in this way until we will find a state $\Psi (m_i,s_{m_i};
\dots;0,0)$ for which we have $h \geq s_{m_i}$ (it will always exist, as  $s_{0}=0$).\\
Then, using equations $\Delta^{(h)}(F^-)\Psi (m_i,s_{m_i},\dots,0,0)=0$ and (\ref{deltah1}) we obtain:
\begin{equation}
C_{h} \Psi (m_i,s_{m_i};\dots;0,0)=(h-m_i)(h-(m_i-1))\Psi (m_i,s_{m_i};\dots;0,0)\label{eigenvalue5}
\end{equation}
Summarizing:
\begin{equation}
C_h \Phi(k;m_l,s_{m_l};\dots;0,0)=(h-m_i)(h-(m_i-1))\Phi(k;m_l,s_{m_l};\dots;0,0) \label{eigenvalue6}
\end{equation}
where the value of $i \in \{0,1,\dots,l\}$ is selected by the condition:
\begin{equation}
s_{m_i} \leq h <s_{m_{i+1}} \qquad s_{m_{l+1}}=N+1 \label{neq}
\end{equation}

\renewcommand{\thesection}{\Roman{section}}
\section{ A more general Hamiltonian}
\setcounter{equation}{0}
\renewcommand{\thesection}{\arabic{section}}
The basis that  we have constructed in section IV diagonalizes also the more general Hamiltonian $\mathcal{H}$:
\begin{equation}
\mathcal{H}=\lambda \Delta^{(N)}(C_b)+ \mu \Delta^{(N)}(C_f)
\end{equation}
which is an arbitrary linear combination of the bosonic and fermionic part of the Gaudin Hamiltonian $\mathcal{H}_{G_s}$, defined in  (\ref{hgi}).
\begin{equation} 
\Delta^{(N)}(C)=\Delta^{(N)}(C_b)+\Delta^{(N)}(C_f)
\end{equation}
\begin{equation}
C_b=H^2+2(E^+E^-+E^-E^+)
\end{equation}
\begin{equation}
C_f=F^-F^+-F^+F^-
\end{equation}
Indeed, $\mathcal{H}$ can be written as 
\begin{equation}
\mathcal{H}=\mu \Delta^{(N)}(C)+ (\lambda -\mu )\Delta^{(N)}(C_b)
\end{equation}
which obviously commutes with $\Delta^{(h)}(C)$, ($h=2,\dots ,N$) and $\Delta^{(N)}(H)$. So the 
complete set of observables
\begin{equation}
\{\Delta^{(N)}(H),C_2,\dots,C_{N-1} ,\mathcal{H}\}
\end{equation}
has the same common eigenstates as the set of observables (\ref{osser}). Obviously, the eigenvalues of the partial Casimir operators $C_h$ and of $\Delta^{(N)}(H)$ do not change, being still given by (\ref{eigenvalue1}-\ref{eigenvalue6}), while the eigenvalues of the Hamiltonian $\mathcal{H}$  read:
\begin{eqnarray}
&&\mathcal{H} \Phi(k;m_l,s_{m-l};\dots;0,0)=\Big(\lambda(N-m_l)\big(N-(m_l-2)\big)-\mu(N-m_l) +
\nonumber\\
{}&&+ (\mu - \lambda)(1 - (-1)^{k-m_l})(N-m_l + \frac{1}{2})\Big)\Phi(k;m_l,s_{m_l};\dots;0,0)
\nonumber\\
{}&&\label{eigen}
\end{eqnarray}

\renewcommand{\thesection}{\Roman{section}}
\section{Concluding remarks}
\setcounter{equation}{0}
\renewcommand{\thesection}{\arabic{section}}
For the sake of simplicity, the explicit evaluation of the eigenstates has been worked out only  in the $j=\frac{1}{2}$ case. The procedure can be easily 
extended to the case of higher $j$ in each site, but then  the single-particle spaces become $4j+1$-dimensional and calculations get much more lenghty.\\
A natural generalization of the model presented here will be associated with the $q$-deformation of the superalgebra $osp(1,2)$ starting from the results 
first derived by Kulish already in the late $80$'s, and using the later findings by Lukierski ~\cite{lukierski}~\cite{kuli}~\cite{kul}. In the coproduct approach integrability of such $q$-deformed supersymmetric models will follow from the Hopf-algebra structure of $U_q(osp(1,2))$. A further extension is related to the investigation of superalgebras other than $osp(1,2)$, for instance $sl(1,2)$, already introduced within the Bethe-Ansatz approach~\cite{frappat}~\cite{gould}. 
Work is progressing in these directions.

\renewcommand{\thesection}{\Roman{section}}
\section{Appendix} 
\setcounter{equation}{0}
\renewcommand{\thesection}{\arabic{section}}
Here we shall prove that our procedure allows one to find all the lowest weight vectors of the representation $(D_{\frac{1}{2}})^N$ (\footnote{ Representation of the chain of $N$ sites with spin $\frac{1}{2}$ in each site.}). These lowest weight vectors are the eigenvectors spanning Ker$\Delta^{(N)}(F^-)$. In fact the construction of all the common eigenvectors of the set of observables (\ref{osser}) consists of two steps. First, we find the `lowest weight vectors'  $\Psi(m_l,s_{m_l},\dots ,0,0)$, built up  on the `total' pseudovacuum' $\Psi (0,0)$. Then we construct all the remaining states by applying $\Delta^{(N)}(F^+)$ repeatedly on $\Psi(m_l,s_{m_l},\dots ,0,0)$ (\footnote{The number of times we can apply $\Delta^{(N)}(F^+)$ on the state $\Psi(m_l,s_{m_l},\dots ,0,0)$ basically depends on the excitation-number $m_l$ and on the magnetization-number $k$.}).\\
First of all we notice that it holds:
\begin{equation}
(D_{\frac{1}{2}})^N=(D_{\frac{1}{2}})^{(N-1)}\otimes D_{\frac{1}{2}}\label{gordan}
\end{equation}
In general the dimension of Ker$\Delta^{(N)}(F^-)$ is obtained recurrently from the Clebsch-Gordan series:
\begin{equation}
D_j\otimes D_k=D_{j+k}\oplus \dots \oplus D_{|j-k|}\label{gor}
\end{equation}
where  $D_{j+k},\dots, D_{|j-k|}$ are irreducible representations, each one possessing one lowest weight vector. Thus, to the representation $(D_j)^N$ it is associated a number of lowest weight vectors, which corresponds to the number of irreducible representations appearing in its decomposition.\\
We have:
\begin{equation}
(D_j)^N=\sum_{k=0}^{Nj}c_{j,k}^{(N)}D_k,\qquad k=0,\frac{1}{2},1,\dots \label{series}
\end{equation}
where $c_{j,k}^{(N)}$ are the appropriate Clebsch-Gordan coefficients.\\
For $j=\frac{1}{2}$ (\ref{series}) obviously becomes:
\begin{equation}
(D_{\frac{1}{2}})^N=\sum_{k=0}^{\frac{N}{2}}c_{{\frac{1}{2}},k}^{(N)}D_k,\qquad k=0,\frac{1}{2},1,\dots \label{serie2}
\end{equation}
The dimension of Ker$\Delta^{(N)}(F^-)$ is given by $\sum_{k=0}^{\frac{N}{2}}c_{{\frac{1}{2}},k}^{(N)}$. On the other hand (\ref{serie2}) can be written in the form (\ref{gordan}):
\begin{equation}
(D_{\frac{1}{2}})^{N}=(\sum_{k=0}^{\frac{N-1}{2}}c_{{\frac{1}{2}},k}^{(N-1)}D_k)\otimes D_{\frac{1}{2}}=\sum_{k=0}^{\frac{N-1}{2}}c_{{\frac{1}{2}},k}^{(N-1)}(D_k\otimes D_{\frac{1}{2}})\label{representation}
\end{equation}
Thus, known all the lowest weight vectors of the representation $(D_{\frac{1}{2}})^{N-1}$, by induction we have all the lowest weight vectors of the representation $(D_{\frac{1}{2}})^N$. There are two different cases:
\begin{equation}
D_k\otimes D_{\frac{1}{2}}= D_{\frac{1}{2}}\qquad k=0\label{gord}
\end{equation}
There is only one lowest weight vector.
\begin{equation}
D_k\otimes D_{\frac{1}{2}}=D_{k-\frac{1}{2}}\oplus D_k \oplus D_{k+\frac{1}{2}}\qquad k\geq \frac{1}{2}\label{gorda}
\end{equation}
We have $3$ lowest weight vectors.\\
So from the Clebsh-Gordan series we learn that the dimension of Ker$\Delta^{(N)}(F^-)$ can be obtained recurrently from the dimension of  Ker$\Delta^{(N-1)}(F^-)$, observing that a generic state of the kernel of $\Delta^{(N-1)}(F^-)$ will give rise to one new state of the kernel of $\Delta^{(N)}(F^-)$ for $k=0$ and to three new states otherwise ($k\neq 0$). \\
Has our procedure the same starting points and the same properties?\\
To both questions the answer is affirmative. First, concerning the starting point ($N=2$), we notice that formula (\ref{gordan}) yields:
\begin{equation}
(D_{\frac{1}{2}})^2=D_0\oplus D_{\frac{1}{2}}\oplus D_1\label{dunmezzo}
\end{equation}
while with our technique from (\ref{cond},\ref{states2},\ref{vacuum}) we get:
\begin{eqnarray}
&&\Psi(0,0)=\bigg( \hspace{-1.2mm}\begin{array}{c} \scriptstyle{0} \vspace{-5mm}\\ \scriptstyle{1}\vspace{-5mm}\\ \scriptstyle{0} \end{array} \hspace{-1.2mm} \bigg)\bigg( \hspace{-1.2mm}\begin{array}{c} \scriptstyle{0} \vspace{-5mm}\\ \scriptstyle{1}\vspace{-5mm}\\ \scriptstyle{0} \end{array} \hspace{-1.2mm} \bigg)\nonumber\\
{}&&\Psi(1,2;0,0)=\sum_{i=0}^1 a_i (F_1^+)^{(1-i)}(F_2^+)^i \Psi(0,0)=\nonumber\\
{}&&=a_0 F_1^+\bigg( \hspace{-1.2mm}\begin{array}{c} \scriptstyle{0} \vspace{-5mm}\\ \scriptstyle{1}\vspace{-5mm}\\ \scriptstyle{0} \end{array} \hspace{-1.2mm} \bigg)\bigg( \hspace{-1.2mm}\begin{array}{c} \scriptstyle{0} \vspace{-5mm}\\ \scriptstyle{1}\vspace{-5mm}\\ \scriptstyle{0} \end{array} \hspace{-1.2mm} \bigg)+a_1 F_2^+ \bigg( \hspace{-1.2mm}\begin{array}{c} \scriptstyle{0} \vspace{-5mm}\\ \scriptstyle{1}\vspace{-5mm}\\ \scriptstyle{0} \end{array} \hspace{-1.2mm} \bigg)\bigg( \hspace{-1.2mm}\begin{array}{c} \scriptstyle{0} \vspace{-5mm}\\ \scriptstyle{1}\vspace{-5mm}\\ \scriptstyle{0} \end{array} \hspace{-1.2mm} \bigg)\nonumber\\
{}&&\Psi(2,2;0,0,0)=\sum_{i=0}^2 a_i (F_1^+)^{(2-i)}(F_2^+)^i \Psi(0,0)=\nonumber\\
{}&&=a_0 (F_1^+)^2\bigg( \hspace{-1.2mm}\begin{array}{c} \scriptstyle{0} \vspace{-5mm}\\ \scriptstyle{1}\vspace{-5mm}\\ \scriptstyle{0} \end{array} \hspace{-1.2mm} \bigg)\bigg( \hspace{-1.2mm}\begin{array}{c} \scriptstyle{0} \vspace{-5mm}\\ \scriptstyle{1}\vspace{-5mm}\\ \scriptstyle{0} \end{array} \hspace{-1.2mm} \bigg)+ a_1F_1^+ F_2^+ \bigg( \hspace{-1.2mm}\begin{array}{c} \scriptstyle{0} \vspace{-5mm}\\ \scriptstyle{1}\vspace{-5mm}\\ \scriptstyle{0} \end{array} \hspace{-1.2mm} \bigg)\bigg( \hspace{-1.2mm}\begin{array}{c} \scriptstyle{0} \vspace{-5mm}\\ \scriptstyle{1}\vspace{-5mm}\\ \scriptstyle{0} \end{array} \hspace{-1.2mm} \bigg)+a_2 (F_2^+)^2\bigg( \hspace{-1.2mm}\begin{array}{c} \scriptstyle{0} \vspace{-5mm}\\ \scriptstyle{1}\vspace{-5mm}\\ \scriptstyle{0} \end{array} \hspace{-1.2mm} \bigg)\bigg( \hspace{-1.2mm}\begin{array}{c} \scriptstyle{0} \vspace{-5mm}\\ \scriptstyle{1}\vspace{-5mm}\\ \scriptstyle{0} \end{array} \hspace{-1.2mm} \bigg)\nonumber\\
{}&&
\end{eqnarray}
and from the conditions on the coefficients $a_i$ (\ref{condone},\ref{condtwo1},\ref{condtwo2}) we get:
\begin{eqnarray*}
&&\Psi(1,2;0,0)=\bigg( \hspace{-1.2mm} \begin{array}{c} \scriptstyle{0} \vspace{-5mm}\\ \scriptstyle{0}\vspace{-5mm}\\ \scriptstyle{1} \end{array} \hspace{-1.2mm} \bigg)\bigg( \hspace{-1.2mm} \begin{array}{c} \scriptstyle{0} \vspace{-5mm}\\ \scriptstyle{1}\vspace{-5mm}\\ \scriptstyle{0} \end{array} \hspace{-1.2mm} \bigg) - \bigg( \hspace{-1.2mm} \begin{array}{c} \scriptstyle{0} \vspace{-5mm}\\ \scriptstyle{1}\vspace{-5mm}\\ \scriptstyle{0} \end{array} \hspace{-1.2mm} \bigg)\bigg( \hspace{-1.2mm} \begin{array}{c} \scriptstyle{0} \vspace{-5mm}\\ \scriptstyle{0}\vspace{-5mm}\\ \scriptstyle{1} \end{array} \hspace{-1.2mm} \bigg)\\
&&\Psi(2,2;0,0)=\bigg( \hspace{-1.2mm} \begin{array}{c} \scriptstyle{1} \vspace{-5mm}\\ \scriptstyle{0}\vspace{-5mm}\\ \scriptstyle{0} \end{array} \hspace{-1.2mm} \bigg)\bigg( \hspace{-1.2mm} \begin{array}{c} \scriptstyle{0} \vspace{-5mm}\\ \scriptstyle{1}\vspace{-5mm}\\ \scriptstyle{0} \end{array} \hspace{-1.2mm} \bigg) - \bigg( \hspace{-1.2mm} \begin{array}{c} \scriptstyle{0} \vspace{-5mm}\\ \scriptstyle{0}\vspace{-5mm}\\ \scriptstyle{1} \end{array} \hspace{-1.2mm} \bigg)\bigg( \hspace{-1.2mm} \begin{array}{c} \scriptstyle{0} \vspace{-5mm}\\ \scriptstyle{0}\vspace{-5mm}\\ \scriptstyle{1} \end{array} \hspace{-1.2mm} \bigg) - \bigg( \hspace{-1.2mm} \begin{array}{c} \scriptstyle{0} \vspace{-5mm}\\ \scriptstyle{1}\vspace{-5mm}\\ \scriptstyle{0} \end{array} \hspace{-1.2mm} \bigg)\bigg( \hspace{-1.2mm} \begin{array}{c} \scriptstyle{1} \vspace{-5mm}\\ \scriptstyle{0}\vspace{-5mm}\\ \scriptstyle{0} \end{array} \hspace{-1.2mm} \bigg)
\end{eqnarray*}
So we have $3$ states according to (\ref{dunmezzo}).\\
Now we turn to consider the general case. Let $\psi ^{(N-1)}(m_l, s_{m_l},\dots,0,0)$ be a state belonging to Ker$(\Delta ^{(N-1)} (F^-))$: we will obtain a state belonging to  Ker$(\Delta ^{(N)} (F^-))$ by tensorizing it with a ``spin down'' state:
$$\psi ^{(N)}(m_l, s_{m_l},\dots,0,0)~= \psi ^{(N-1)}(m_l, s_{m_l},\dots,0,0) \otimes   
\bigg( \hspace{-1.2mm}\begin{array}{c} \scriptstyle{0} \vspace{-5mm}\\ \scriptstyle{1}\vspace{-5mm}\\ \scriptstyle{0} \end{array} \hspace{-1.2mm} \bigg)$$

\noindent
As $\psi ^{(N-1)}(m_l, s_{m_l},\dots,0,0)$ spans Ker$(\Delta ^{(N-1)} (F^-))$ we get all states
belonging to Ker$(\Delta ^{(N)} (F^-))$ having $s_{m_l}<N$. When  generating the complementary subspace of Ker$(\Delta ^{(N)} (F^-))$ through our algorithm, 
we have to consider two different cases:
\begin{enumerate}
\item  $\psi ^{(N-1)}(m_l, s_{m_l},\dots,0,0)$ is annihilated by $\Delta ^{(N-1)}(H)$. In this case it holds $m_l - N +1 = 0$ $~\Rightarrow~$ no further element of Ker$(\Delta ^{(N)} (F^-))$ can be constructed, as, from (\ref{ine}), it would have $s_{m_{l+1}}~>~N$.
\item  $\psi ^{(N-1)}(m_l, s_{m_l},\dots,0,0)$ does not belong to Ker$\Delta ^{(N-1)}(H)$. it follows that $m_l \le N-2$, 
and consequently two more states of Ker$(\Delta ^{(N)} (F^-))$ can be costructed, with \\ $(m_{l+1}, s_{m_{l+1}}) = (m_l+1,N)$ and $(m_{l+1}, s_{m_{l+1}}) = (m_l+2,N)$ 
respectively.
\end{enumerate}
Summarizing, we obtain one element of Ker$(\Delta ^{(N)} (F^-))$ starting from each  $(N-1)$-particle state of spin $0$, and three elements of 
Ker$(\Delta ^{(N)} (F^-))$ starting from each  $(N-1)$-particle state of spin $k>0$.

\medskip
\noindent
As an example, we give the explicit form of the states (\ref{states2}) in the case $N=3$ and for the choice of the normalization parameter $a_0=1$. This is the basis spanning the kernel of $\Delta^{(N)}(F^-)$:

\begin{eqnarray*} 
\Psi(0,0)&=&\bigg( \hspace{-1.2mm}\begin{array}{c} \scriptstyle{0} \vspace{-5mm}\\ \scriptstyle{1}\vspace{-5mm}\\ \scriptstyle{0} \end{array} \hspace{-1.2mm} \bigg)\bigg( \hspace{-1.2mm}\begin{array}{c} \scriptstyle{0} \vspace{-5mm}\\ \scriptstyle{1}\vspace{-5mm}\\ \scriptstyle{0} \end{array} \hspace{-1.2mm} \bigg)\bigg( \hspace{-1.2mm} \begin{array}{c} \scriptstyle{0} \vspace{-5mm}\\ \scriptstyle{1}\vspace{-5mm}\\ \scriptstyle{0} \end{array} \hspace{-1.2mm} \bigg)\\
\Psi(1,2;0,0)&=&\bigg( \hspace{-1.2mm} \begin{array}{c} \scriptstyle{0} \vspace{-5mm}\\ \scriptstyle{0}\vspace{-5mm}\\ \scriptstyle{1} \end{array} \hspace{-1.2mm} \bigg)\bigg( \hspace{-1.2mm} \begin{array}{c} \scriptstyle{0} \vspace{-5mm}\\ \scriptstyle{1}\vspace{-5mm}\\ \scriptstyle{0} \end{array} \hspace{-1.2mm} \bigg)\bigg( \hspace{-1.2mm} \begin{array}{c} \scriptstyle{0} \vspace{-5mm}\\ \scriptstyle{1}\vspace{-5mm}\\ \scriptstyle{0} \end{array} \hspace{-1.2mm} \bigg) - \bigg( \hspace{-1.2mm} \begin{array}{c} \scriptstyle{0} \vspace{-5mm}\\ \scriptstyle{1}\vspace{-5mm}\\ \scriptstyle{0} \end{array} \hspace{-1.2mm} \bigg)\bigg( \hspace{-1.2mm} \begin{array}{c} \scriptstyle{0} \vspace{-5mm}\\ \scriptstyle{0}\vspace{-5mm}\\ \scriptstyle{1} \end{array} \hspace{-1.2mm} \bigg)\bigg( \begin{array}{c} \hspace{-1.2mm} \scriptstyle{0} \vspace{-5mm}\\ \scriptstyle{1}\vspace{-5mm}\\ \scriptstyle{0} \end{array} \hspace{-1.2mm} \bigg)\\
\Psi(1,3;0,0)&=&\bigg( \hspace{-1.2mm} \begin{array}{c} \scriptstyle{0} \vspace{-5mm}\\ \scriptstyle{0}\vspace{-5mm}\\ \scriptstyle{1} \end{array} \hspace{-1.2mm} \bigg)\bigg( \hspace{-1.2mm} \begin{array}{c} \scriptstyle{0} \vspace{-5mm}\\ \scriptstyle{1}\vspace{-5mm}\\ \scriptstyle{0} \end{array} \hspace{-1.2mm} \bigg)\bigg( \hspace{-1.2mm} \begin{array}{c} \scriptstyle{0} \vspace{-5mm}\\ \scriptstyle{1}\vspace{-5mm}\\ \scriptstyle{0} \end{array} \hspace{-1.2mm} \bigg) + \bigg( \hspace{-1.2mm} \begin{array}{c} \scriptstyle{0} \vspace{-5mm}\\ \scriptstyle{1}\vspace{-5mm}\\ \scriptstyle{0} \end{array} \hspace{-1.2mm} \bigg)\bigg( \hspace{-1.2mm} \begin{array}{c} \scriptstyle{0} \vspace{-5mm}\\ \scriptstyle{0}\vspace{-5mm}\\ \scriptstyle{1} \end{array} \hspace{-1.2mm} \bigg)\bigg( \hspace{-1.2mm} \begin{array}{c} \scriptstyle{0} \vspace{-5mm}\\ \scriptstyle{1}\vspace{-5mm}\\ \scriptstyle{0} \end{array} \hspace{-1.2mm} \bigg) - 2\bigg( \hspace{-1.2mm} \begin{array}{c} \scriptstyle{0} \vspace{-5mm}\\ \scriptstyle{1}\vspace{-5mm}\\ \scriptstyle{0} \end{array} \hspace{-1.2mm} \bigg)\bigg( \hspace{-1.2mm} \begin{array}{c} \scriptstyle{0} \vspace{-5mm}\\ \scriptstyle{1}\vspace{-5mm}\\ \scriptstyle{0} \end{array} \hspace{-1.2mm} \bigg)\bigg( \hspace{-1.2mm} \begin{array}{c} \scriptstyle{0} \vspace{-5mm}\\ \scriptstyle{0}\vspace{-5mm}\\ \scriptstyle{1} \end{array} \hspace{-1.2mm} \bigg)
\end{eqnarray*}

\begin{eqnarray*}
\Psi(2,2;0,0)&=&\bigg( \hspace{-1.2mm} \begin{array}{c} \scriptstyle{1} \vspace{-5mm}\\ \scriptstyle{0}\vspace{-5mm}\\ \scriptstyle{0} \end{array} \hspace{-1.2mm} \bigg)\bigg( \hspace{-1.2mm} \begin{array}{c} \scriptstyle{0} \vspace{-5mm}\\ \scriptstyle{1}\vspace{-5mm}\\ \scriptstyle{0} \end{array} \hspace{-1.2mm} \bigg)\bigg( \hspace{-1.2mm} \begin{array}{c} \scriptstyle{0} \vspace{-5mm}\\ \scriptstyle{1}\vspace{-5mm}\\ \scriptstyle{0} \end{array} \hspace{-1.2mm}\bigg) - \bigg( \hspace{-1.2mm} \begin{array}{c} \scriptstyle{0} \vspace{-5mm}\\ \scriptstyle{0}\vspace{-5mm}\\ \scriptstyle{1} \end{array} \hspace{-1.2mm} \bigg)\bigg( \hspace{-1.2mm} \begin{array}{c} \scriptstyle{0} \vspace{-5mm}\\ \scriptstyle{0}\vspace{-5mm}\\ \scriptstyle{1} \end{array} \hspace{-1.2mm} \bigg)\bigg( \hspace{-1.2mm} \begin{array}{c} \scriptstyle{0} \vspace{-5mm}\\ \scriptstyle{1}\vspace{-5mm}\\ \scriptstyle{0} \end{array} \hspace{-1.2mm} \bigg) - \bigg( \hspace{-1.2mm} \begin{array}{c} \scriptstyle{0} \vspace{-5mm}\\ \scriptstyle{1}\vspace{-5mm}\\ \scriptstyle{0} \end{array} \hspace{-1.2mm} \bigg)\bigg( \hspace{-1.2mm} \begin{array}{c} \scriptstyle{1} \vspace{-5mm}\\ \scriptstyle{0}\vspace{-5mm}\\ \scriptstyle{0} \end{array} \hspace{-1.2mm} \bigg)\bigg( \hspace{-1.2mm} \begin{array}{c} \scriptstyle{0} \vspace{-5mm}\\ \scriptstyle{1}\vspace{-5mm}\\ \scriptstyle{0} \end{array} \hspace{-1.2mm} \bigg)
\end{eqnarray*}
\begin{eqnarray*}
\Psi(2,3;0,0)&=&\bigg( \hspace{-1.2mm}\begin{array}{c} \scriptstyle{1} \vspace{-5mm}\\ \scriptstyle{0}\vspace{-5mm}\\ \scriptstyle{0} \end{array} \hspace{-1.2mm} \bigg)\bigg( \hspace{-1.2mm}\begin{array}{c} \scriptstyle{0} \vspace{-5mm}\\ \scriptstyle{1}\vspace{-5mm}\\ \scriptstyle{0} \end{array} \hspace{-1.2mm} \bigg)\bigg( \hspace{-1.2mm}\begin{array}{c} \scriptstyle{0} \vspace{-5mm}\\ \scriptstyle{1}\vspace{-5mm}\\ \scriptstyle{0} \end{array} \hspace{-1.2mm} \bigg) + \bigg( \hspace{-1.2mm}\begin{array}{c} \scriptstyle{0} \vspace{-5mm}\\ \scriptstyle{1}\vspace{-5mm}\\ \scriptstyle{0} \end{array} \hspace{-1.2mm} \bigg)\bigg( \hspace{-1.2mm}\begin{array}{c} \scriptstyle{1} \vspace{-5mm}\\ \scriptstyle{0}\vspace{-5mm}\\ \scriptstyle{0} \end{array} \hspace{-1.2mm} \bigg)\bigg( \hspace{-1.2mm}\begin{array}{c} \scriptstyle{0} \vspace{-5mm}\\ \scriptstyle{1}\vspace{-5mm}\\ \scriptstyle{0} \end{array} \hspace{-1.2mm} \bigg) - \bigg( \hspace{-1.2mm}\begin{array}{c} \scriptstyle{0} \vspace{-5mm}\\ \scriptstyle{0}\vspace{-5mm}\\ \scriptstyle{1} \end{array} \hspace{-1.2mm} \bigg)\bigg( \hspace{-1.2mm}\begin{array}{c} \scriptstyle{0} \vspace{-5mm}\\ \scriptstyle{1}\vspace{-5mm}\\ \scriptstyle{0} \end{array} \hspace{-1.2mm} \bigg)\bigg( \hspace{-1.2mm}\begin{array}{c} \scriptstyle{0} \vspace{-5mm}\\ \scriptstyle{0}\vspace{-5mm}\\ \scriptstyle{1} \end{array} \hspace{-1.2mm} \bigg) -\\
&-& \bigg( \hspace{-1.2mm}\begin{array}{c} \scriptstyle{0} \vspace{-5mm}\\ \scriptstyle{1}\vspace{-5mm}\\ \scriptstyle{0} \end{array} \hspace{-1.2mm} \bigg)\bigg( \hspace{-1.2mm}\begin{array}{c} \scriptstyle{0} \vspace{-5mm}\\ \scriptstyle{0}\vspace{-5mm}\\ \scriptstyle{1} \end{array} \hspace{-1.2mm} \bigg)\bigg( \hspace{-1.2mm}\begin{array}{c} \scriptstyle{0} \vspace{-5mm}\\ \scriptstyle{0}\vspace{-5mm}\\ \scriptstyle{1} \end{array} \hspace{-1.2mm} \bigg) -2\bigg( \hspace{-1.2mm}\begin{array}{c} \scriptstyle{0} \vspace{-5mm}\\ \scriptstyle{1}\vspace{-5mm}\\ \scriptstyle{0} \end{array} \hspace{-1.2mm} \bigg)\bigg( \hspace{-1.2mm}\begin{array}{c} \scriptstyle{0} \vspace{-5mm}\\ \scriptstyle{1}\vspace{-5mm}\\ \scriptstyle{0} \end{array} \hspace{-1.2mm} \bigg)\bigg( \hspace{-1.2mm}\begin{array}{c} \scriptstyle{1} \vspace{-5mm}\\ \scriptstyle{0}\vspace{-5mm}\\ \scriptstyle{0} \end{array} \hspace{-1.2mm} \bigg)
\end{eqnarray*}

\begin{eqnarray*}
\Psi(2,3;1,2;0,0)&=&\bigg( \hspace{-1.2mm} \begin{array}{c} \scriptstyle{1} \vspace{-5mm}\\ \scriptstyle{0}\vspace{-5mm}\\ \scriptstyle{0} \end{array} \hspace{-1.2mm} \bigg)\bigg( \hspace{-1.2mm} \begin{array}{c} \scriptstyle{0} \vspace{-5mm}\\ \scriptstyle{1}\vspace{-5mm}\\ \scriptstyle{0} \end{array} \hspace{-1.2mm} \bigg)\bigg( \hspace{-1.2mm} \begin{array}{c} \scriptstyle{0} \vspace{-5mm}\\ \scriptstyle{1}\vspace{-5mm}\\ \scriptstyle{0} \end{array} \hspace{-1.2mm} \bigg) - 2\bigg( \hspace{-1.2mm} \begin{array}{c} \scriptstyle{0} \vspace{-5mm}\\ \scriptstyle{0}\vspace{-5mm}\\ \scriptstyle{1} \end{array} \hspace{-1.2mm} \bigg)\bigg( \hspace{-1.2mm} \begin{array}{c} \scriptstyle{0} \vspace{-5mm}\\ \scriptstyle{0}\vspace{-5mm}\\ \scriptstyle{1} \end{array} \hspace{-1.2mm} \bigg)\bigg( \hspace{-1.2mm} \begin{array}{c} \scriptstyle{0} \vspace{-5mm}\\ \scriptstyle{1}\vspace{-5mm}\\ \scriptstyle{0} \end{array} \hspace{-1.2mm} \bigg) - \bigg( \hspace{-1.2mm} \begin{array}{c} \scriptstyle{0} \vspace{-5mm}\\ \scriptstyle{1}\vspace{-5mm}\\ \scriptstyle{0} \end{array} \hspace{-1.2mm} \bigg)\bigg( \hspace{-1.2mm} \begin{array}{c} \scriptstyle{1} \vspace{-5mm}\\ \scriptstyle{0}\vspace{-5mm}\\ \scriptstyle{0} \end{array} \hspace{-1.2mm} \bigg)\bigg( \hspace{-1.2mm} \begin{array}{c} \scriptstyle{0} \vspace{-5mm}\\ \scriptstyle{1}\vspace{-5mm}\\ \scriptstyle{0} \end{array} \hspace{-1.2mm} \bigg)-\\
&-&\bigg( \hspace{-1.2mm}\begin{array}{c} \scriptstyle{0} \vspace{-5mm}\\ \scriptstyle{0}\vspace{-5mm}\\ \scriptstyle{1} \end{array} \hspace{-1.2mm} \bigg)\bigg( \hspace{-1.2mm}\begin{array}{c} \scriptstyle{0} \vspace{-5mm}\\ \scriptstyle{1}\vspace{-5mm}\\ \scriptstyle{0} \end{array} \hspace{-1.2mm} \bigg)\bigg( \hspace{-1.2mm} \begin{array}{c} \scriptstyle{0} \vspace{-5mm}\\ \scriptstyle{0}\vspace{-5mm}\\ \scriptstyle{1} \end{array} \hspace{-1.2mm} \bigg) + \bigg( \hspace{-1.2mm}\begin{array}{c} \scriptstyle{0} \vspace{-5mm}\\ \scriptstyle{1}\vspace{-5mm}\\ \scriptstyle{0} \end{array} \hspace{-1.2mm} \bigg)\bigg( \hspace{-1.2mm}\begin{array}{c} \scriptstyle{0} \vspace{-5mm}\\ \scriptstyle{0}\vspace{-5mm}\\ \scriptstyle{1} \end{array} \hspace{-1.2mm} \bigg)\bigg( \hspace{-1.2mm} \begin{array}{c} \scriptstyle{0} \vspace{-5mm}\\ \scriptstyle{0}\vspace{-5mm}\\ \scriptstyle{1} \end{array} \hspace{-1.2mm} \bigg)
\end{eqnarray*}
\begin{eqnarray*}
\Psi(3,3;1,2;0,0)&=&\bigg( \hspace{-1.2mm}\begin{array}{c} \scriptstyle{0} \vspace{-5mm}\\ \scriptstyle{0}\vspace{-5mm}\\ \scriptstyle{1} \end{array} \hspace{-1.2mm} \bigg)\bigg( \hspace{-1.2mm}\begin{array}{c} \scriptstyle{1} \vspace{-5mm}\\ \scriptstyle{0}\vspace{-5mm}\\ \scriptstyle{0} \end{array} \hspace{-1.2mm} \bigg)\bigg( \hspace{-1.2mm} \begin{array}{c} \scriptstyle{0} \vspace{-5mm}\\ \scriptstyle{1}\vspace{-5mm}\\ \scriptstyle{0} \end{array} \hspace{-1.2mm} \bigg) - \bigg( \hspace{-1.2mm}\begin{array}{c} \scriptstyle{0} \vspace{-5mm}\\ \scriptstyle{0}\vspace{-5mm}\\ \scriptstyle{1} \end{array} \hspace{-1.2mm} \bigg)\bigg( \hspace{-1.2mm}\begin{array}{c} \scriptstyle{0} \vspace{-5mm}\\ \scriptstyle{1}\vspace{-5mm}\\ \scriptstyle{0} \end{array} \hspace{-1.2mm} \bigg)\bigg( \hspace{-1.2mm} \begin{array}{c} \scriptstyle{1} \vspace{-5mm}\\ \scriptstyle{0}\vspace{-5mm}\\ \scriptstyle{0} \end{array} \hspace{-1.2mm} \bigg) - \bigg( \hspace{-1.2mm}\begin{array}{c} \scriptstyle{1} \vspace{-5mm}\\ \scriptstyle{0}\vspace{-5mm}\\ \scriptstyle{0} \end{array} \hspace{-1.2mm} \bigg)\bigg( \hspace{-1.2mm}\begin{array}{c} \scriptstyle{0} \vspace{-5mm}\\ \scriptstyle{0}\vspace{-5mm}\\ \scriptstyle{1} \end{array} \hspace{-1.2mm} \bigg)\bigg( \hspace{-1.2mm} \begin{array}{c} \scriptstyle{0} \vspace{-5mm}\\ \scriptstyle{1}\vspace{-5mm}\\ \scriptstyle{0} \end{array} \hspace{-1.2mm} \bigg) +\\
&+& \bigg( \hspace{-1.2mm}\begin{array}{c} \scriptstyle{0} \vspace{-5mm}\\ \scriptstyle{1}\vspace{-5mm}\\ \scriptstyle{0} \end{array} \hspace{-1.2mm} \bigg)\bigg( \hspace{-1.2mm}\begin{array}{c} \scriptstyle{0} \vspace{-5mm}\\ \scriptstyle{0}\vspace{-5mm}\\ \scriptstyle{1} \end{array} \hspace{-1.2mm} \bigg)\bigg( \hspace{-1.2mm} \begin{array}{c} \scriptstyle{1} \vspace{-5mm}\\ \scriptstyle{0}\vspace{-5mm}\\ \scriptstyle{0} \end{array} \hspace{-1.2mm} \bigg) +  \bigg( \hspace{-1.2mm}\begin{array}{c} \scriptstyle{1} \vspace{-5mm}\\ \scriptstyle{0}\vspace{-5mm}\\ \scriptstyle{0} \end{array} \hspace{-1.2mm} \bigg)\bigg( \hspace{-1.2mm}\begin{array}{c} \scriptstyle{0} \vspace{-5mm}\\ \scriptstyle{1}\vspace{-5mm}\\ \scriptstyle{0} \end{array} \hspace{-1.2mm} \bigg)\bigg( \hspace{-1.2mm} \begin{array}{c} \scriptstyle{0} \vspace{-5mm}\\ \scriptstyle{0}\vspace{-5mm}\\ \scriptstyle{1} \end{array} \hspace{-1.2mm} \bigg) - \bigg( \hspace{-1.2mm}\begin{array}{c} \scriptstyle{0} \vspace{-5mm}\\ \scriptstyle{1}\vspace{-5mm}\\ \scriptstyle{0} \end{array} \hspace{-1.2mm} \bigg)\bigg( \hspace{-1.2mm}\begin{array}{c} \scriptstyle{1} \vspace{-5mm}\\ \scriptstyle{0}\vspace{-5mm}\\ \scriptstyle{0} \end{array} \hspace{-1.2mm} \bigg)\bigg( \hspace{-1.2mm} \begin{array}{c} \scriptstyle{0} \vspace{-5mm}\\ \scriptstyle{0}\vspace{-5mm}\\ \scriptstyle{1} \end{array} \hspace{-1.2mm} \bigg) -\\
&-&2 \bigg( \hspace{-1.2mm}\begin{array}{c} \scriptstyle{0} \vspace{-5mm}\\ \scriptstyle{0}\vspace{-5mm}\\ \scriptstyle{1} \end{array} \hspace{-1.2mm} \bigg)\bigg( \hspace{-1.2mm}\begin{array}{c} \scriptstyle{0} \vspace{-5mm}\\ \scriptstyle{0}\vspace{-5mm}\\ \scriptstyle{1} \end{array} \hspace{-1.2mm} \bigg)\bigg( \hspace{-1.2mm} \begin{array}{c} \scriptstyle{0} \vspace{-5mm}\\ \scriptstyle{0}\vspace{-5mm}\\ \scriptstyle{1} \end{array} \hspace{-1.2mm} \bigg)
\end{eqnarray*}
The remaining $3^3-7$ eigenstates are constructed using formula (\ref{phi}).

\end{document}